# On Surface Synthesis of a Pure and Long Range Ordered Titanium(IV)-Porphyrin Contact Layer on Titanium Dioxide.

*Giacomo Lovat[†,‡,#] Daniel Forrer, [§] Mikel Abadia,[ʃ] Marcos Dominguez, [†,‡] Maurizio Casarin, [¥,§] Celia Rogero,[ʃ] Andrea Vittadini[§]\* and Luca Floreano[†]\**

† CNR-IOM, Laboratorio TASC in Area Science Park, Basovizza S.S.14, Km. 163.5, I-34149 Trieste, Italy.

‡ Department of Physics, University of Trieste, P.za Europa 1, I-34127 Trieste, Italy.

§ CNR-ICMATE and INSTM, via Marzolo 1, I-35131 Padova, Italy.

ʃ Materials Physics Center MPC, Centro de Física de Materiales (CSIC-UPV/EHU) and Donostia International Physics Center (DIPC), E-20018 San Sebastian, Spain.

¥ Dipartimento di Scienze Chimiche, Università di Padova, via Marzolo 1, I-35131 Padova, Italy.

ABSTRACT. We show the possibility of tailoring the molecular arrangement, as well as the chemical and structural modifications, of porphyrins at the monolayer saturation coverage on $TiO_2(110)$ by Synchrotron photoemission, electron diffraction, STM topography and DFT calculations. Free-base tetra-phenyl-porphyrins (2H-TPP) adsorb on the oxygen rows, where they can spontaneously capture two additional hydrogen atoms at their iminic nitrogens (4H-TPP). Both 2H-TPP and 4H-TPP molecules aggregate into a commensurate phase at the saturation coverage of one monolayer. Upon sample heating, a self-metalation reaction sets in at 100°C, yielding full metalation of the saturated monolayer at ~200°C. The Ti atoms are extracted from the substrate and, by simultaneous dehydrogenation of the pyrrolic nitrogen atoms, incorporated into the porphyrin macrocycle, where they remain coordinated to two oxygen atoms underneath. Neither the adsorption geometry (on-bridge, atop the oxygen rows) nor the molecular arrangement change across the self-metalation transition up to 300°C. On one side, the robustness of this saturation phase makes it a promising system for its implementation into applications for photocatalysis and photovoltaic devices. On the other side, the possible manifestation of metal exchange with the very reactive Ti atoms must be taken into account when designing porphyrin-sensitized solar cells since the critical temperature for the onset of self-metalation is very close to the normal operating temperature of photovoltaic devices.

INTRODUCTION

The knowledge of the porphyrin chemical reactions in the interfacial layer with titania substrates is fundamental for multiple technological applications, spanning from photocatalysis[1-3] to dye-sensitized solar cells (DSSCs).[4,5] By assembling porphyrins at solid surfaces we can design and fabricate hybrid systems with properties engineered at the nanoscale.[6] In particular the on-surface modification of metal-free porphyrins (2H-P) is a viable route to achieve chemical and structural control of molecular overlayers. Specifically selected metal atoms can be incorporated into the polypyrrolic macrocycle of 2H-P supported films either from pre-deposited metal clusters[7,8] or by post- growth metal deposition atop the molecular overlayer.[9] Upon thermal treatment, adsorbed porphyrins can even undergo "self-metalation" reactions by picking up substrate atoms from almost any metal surface.[10-13] This implies the possibility of "trans-metalation" processes, where the metal atom initially carried by the porphyrin is replaced by a more reactive substrate atom.

In comparison with the large body of data available for metal surfaces,[14] much less is known about metalation and self-metalation reactions of tetrapyrrole molecules at transition metal oxide (TMO) surfaces. In spite of the expected smaller flexibility of the covalent bonding with respect to the metal one, there is evidence suggesting that metalation on metal oxide surfaces could be even faster than on metals. First of all, we have to consider that the kinetics of metalation on metal surfaces is strongly influenced by the porphyrin-surface H exchange,[15] while it has been demonstrated that porphyrins of various functional terminations are able to pick up H atoms from the $TiO_2$(110) surface already at room temperature.[16] Secondly, self-metalation reactions of 2H-TPP at metal surfaces are favored by the presence of oxygen.[17,18] Schneider et al. demonstrated that 2H-TPP is converted to Mg-TPP when adsorbed on MgO substrates.[19] Incidentally, the driving force of the process has been attributed to the high affinity of step and corner oxygens of MgO nanocubes towards the hydrogens of 2H-TPP, which once more indicates the importance of H exchange between porphyrins and surface.

Among reducible TMOs, the rutile (1x1)-$TiO_2$(110) surface has become an important playground to investigate the interfacial properties of organic semiconductors at an atomistic level.[20] In fact, its large anisotropic corrugation, originated by protruding rows of oxygen bridging ($O_{br}$) atoms, has been found to be an effective template for the oriented growth of several polyconjugated aromatics.[21-24] The conversion of the free-base tetraphenylporphyrin (2H-TPP) to Ni-TPP at the $TiO_2$(110) surface has been reported to take place at room temperature when 2H-TPP is deposited first, whereas a temperature of 550 K is required when Ni is pre-deposited.[25] In any case, the postgrowth metal deposition results in relatively low metalation yield (40-70% depending on temperature treatment) and unwanted clustering of metal atoms that deteriorates the interfacial electronic properties. We have found that self-metalation of free-base porphyrins can take place also on rutile $TiO_2$(110) terraces upon mild annealing.[26] This reaction has been recently confirmed by an independent study, where it is suggested that 2nd layer neutral molecules (2H-TPP) would incorporate titanyl radicals, $TiO^{2+}$, from the substrate before than the hydrogenated ones (4H-TPP) in the contact layer underneath, which would be less reactive to self-metalation.[27] In order to clarify the phase diagram of free-base porphyrins on $TiO_2$(110), we performed a thorough investigation of the multiple chemical reactions of 2H-TPP at the technologically relevant monolayer coverage by measuring in real time high resolution photoemission (XPS) by Synchrotron radiation. We complemented the XPS chemical characterization with surface diffraction and STM measurements, as well as first principle calculations for the adsorbed molecules at different coverage.

METHODS

**Experimental details.** We performed X-ray photoemission spectroscopy and RHEED measurements at the ALOISA beamline[28] of the Elettra Synchrotron (Trieste, Italy), while topographic images of the film surface were collected with an Aahrus microscope (provided by SPECS) at the Centro de Fisica de Materiales (CSIC-UPV, San Sebastian, Spain).[29] All XPS measurements have been performed in transverse magnetic polarization (TM, i.e. close to p-polarization) and normal emission geometry, with the sample kept a grazing angle of 4˚. The photoemission spectra of the N 1s peak (at hν=500 eV) and of the Valence Band (at hν=140 eV) were calibrated to the binding energy of Ti $3p$ = 37.6±0.05 eV, as previously calibrated to the binding energy of Ti $2p_{3/2}$ = 459.1 eV (at hν=650 eV), thus yielding the Defect State peak at 0.9 eV.

In both laboratories, we used boron nitride Knudsen cells for the sublimation of 2H-TPP (purchased from Sigma-Aldrich, purity 99%). Deposition was typically performed at a temperature of 570±10 K, which corresponds to deposition rates in the range of 0.1-0.5 Å/min, as monitored in situ by room temperature quartz microbalances (assuming a 2H-TPP density of 1.26 g/cm$^3$). In the ALOISA apparatus, the monolayer coverage (ML) has been defined by recording the XPS intensity after thermal desorption of a multilayer at 200-250°C, which yields a saturation layer corresponding to a nominal thickness of ~2.5 Å. At the first use of the 2H-TPP powder, we have seldom observed an excess of the pyrrolic component (non-stoichiometric overall intensity of N 1s) in the submonolayer coverage range and at low temperature that vanishes upon mild annealing to ~40-50°C, well before the onset of the self-metalation reaction. At the same time, the intensity of the C 1s peak remains practically unchanged from low temperature (<30˚C) up to 300˚C for any initial coverage. In addition, RHEED and STM indicate that no significant morphological changes take place in the considered temperature range. These evidences are indicative of residual volatile contaminants (small fragments) in the porphyrin powder, which are weakly physisorbed on the bare substrate, but not on the porphyrin layer.

The experiments have been performed and repeated on several different samples purchased from Mateck that were also exchanged between the ALOISA Synchrotron beamline and the STM apparatus. In order to prevent large temperature differences between the top and bottom surface, as well as local inhomogeneity, we employed samples 0.5 mm thick and 8 mm wide. We followed the same preparation protocol for cleaning the surface: sputtering with Ar$^+$ at 0.8-1.0 keV followed by short annealing up to 1000±50 K, which yields dark blue samples. Two thermocouples were employed in the ALOISA sample holder, in direct contact with a 2 mm thick molybdenum spacer (12 mm diameter disk) underneath the sample, which was attached with a silver paste on the Mo disk to improve the temperature homogeneity during heating (both radiative and electron bombardment heating on the bottom surface of the Mo spacer). Before insertion into vacuum, the sample/spacer assembly was cured by slow (two days) heating in air up to ~400°C in order to completely dry the organic solvent of the silver paste. In the ALOISA experimental chamber, the sample heating can be performed simultaneously with the XPS and RHEED acquisition, hence allowing us to monitor in real time the chemical reactions and to single out the different phases. Alternatively, a few XPS measurements were performed at the ALOISA branchline, HASPES, where molecular deposition can be quantitatively monitored in real time by XPS.[30] Both HASPES and ALOISA experimental chambers share the same kind of sample holder, which is also carrying the thermocouples and the heaters, hence guaranteeing the reproducibility of the thermal treatments. In the STM apparatus, the thermocouple is in contact with the molybdenum manipulator just aside the sample holder. This configuration yields a somehow larger uncertainty on the effective surface temperature thus we only focused on the relevant annealing temperature as previously determined at ALOISA/HASPES. Finally, all the STM images were acquired at room temperature using a Colibrì® tip sensor.

**Computational details.** Calculations were performed within a plane-wave pseudopotential framework, using the Quantum-ESPRESSO (QE) suite of codes.[31] Valence orbitals were expanded on a plane-wave basis set with kinetic energy cutoff of 25 Ry, while the cutoff on the augmentation density was 200 Ry. The PBE-D2 exchange-correlation functional was adopted, where van der Waals interactions are included by means of the Grimme method,[32] as implemented in QE.[33] The interaction between ion cores and valence electrons was modeled by ultrasoft pseudopotentials,[34] whose core include 1s orbitals for C, N and O, and 1s-2p orbitals for Ti. Surfaces were modelled by means of a repeated slab approach with PBE-D2 theoretical lattice constants of the substrate,[35] where the $TiO_2$ substrate is represented by four $TiO_2$ layer, with the bottom two kept frozen in their bulk positions during geometry optimization runs. Due to the size of the surface cells, a Γ-point integration scheme was used.

The adsorption energies of 2H-TPP and 4H-TPP were calculated as difference between the total energy of the adsorbed system and the energy sum of the isolated molecule and the clean $TiO_2$ surface. Specific calculations about the stability of the 4H-TPP molecule in the gas phase and adsorbed configurations can be found in Ref. [16]. The formal adsorption energy of the metalated molecule must be computed using a titanyl TPP (TiO-TPP) as reference. Thus, the adsorbed system is the product of adsorbing such a molecule on an oxygen vacancy of the $TiO_2$ surface. The adsorption energy of the metaled species is defined as the difference between the total energy of the adsorbed system and the energy sum of the isolated molecule and the defective $TiO_2$ surface, plus the vacancy formation energy. The latter is estimated taking as reference an $O_2$ molecule in vacuum, whose total energy was corrected for the GGA overbind error as prescribed by Wang et al.[36] Additional details can be found in our former study.[16] We remark that no comparison is possible between the energetics of the different molecules because the microscopic mechanisms driving the hydrogenation and metalation reactions, hence the initial configurations, are not known. In the present study we rather present comparative energy values between different adsorption geometries and/or different ordered phases for the same molecular species. In this way the energy of the starting configuration for each molecular species is simply a constant offset that is canceled out in the energy balance.

RESULTS AND DISCUSSION
**Porphyrin self-metalation by XPS.** The chemical state of the porphyrin macrocycle can be monitored by measuring the XPS of the N 1s core level. There are two types of nitrogen atoms in free-base 2H-TPP, i.e. iminic nitrogens and pyrrolic nitrogens, whose core levels display separated peaks of equal area at a binding energy (BE) of ~398 eV (iminic) and ~400 eV (pyrrolic). When coordinated to a central metal atom, the four N atoms becomes equivalent, yielding a single N 1s peak, which is shifted by ~0.3-0.5 eV to higher BE with respect to the iminic peak. When 2H-TPP is deposited on the rutile $TiO_2(110)$ surface, both pyrrolic and iminic peaks are shifted to higher binding energy by ~0.3-0.4 eV in the first layer. Most importantly, porphyrins have been shown to capture H atoms from the hydroxyl groups present on the surface already at room temperature (RT), as witnessed by the almost complete vanishing of the iminic peak.[16] These H atoms come from both hydroxyl groups present on the surface at room temperature (RT) and from subsurface buried H. The latter ones are typically associated with metal impurities (not detectable at XPS) in the original samples, eventually in a larger concentration near the surface due to the preliminary thermal treatment in air for drying the silver paste used to attach the samples that may favor either native hydrogen diffusion from the bulk to the subsurface layers or, vice-versa, $TiO_2$ hydrogenation in ambient conditions (specially at defects, and sample edges).[37] The surface hydroxyls are formed by dissociation of residual water at the oxygen vacancies that are left on the surface after high temperature annealing in vacuum,[38] and they are associated with a specific XPS peak in the valence band at BE~11 eV,[39] see upper panel of Fig. 1. Porphyrins effectively capture also H atoms present in the nearby subsurface layers, as can be seen by the comparison of Fig. 1, where the deposition of 0.9 ML of porphyrins on a hydroxyl-free surface yields only a small increase of the residual iminic component with respect to deposition on a surface with 4-5% concentration of hydroxyls. In this regard, we remark that any iminic contribution from 2nd layer molecules would show up at much lower (-0.5 eV) binding energy.[16]

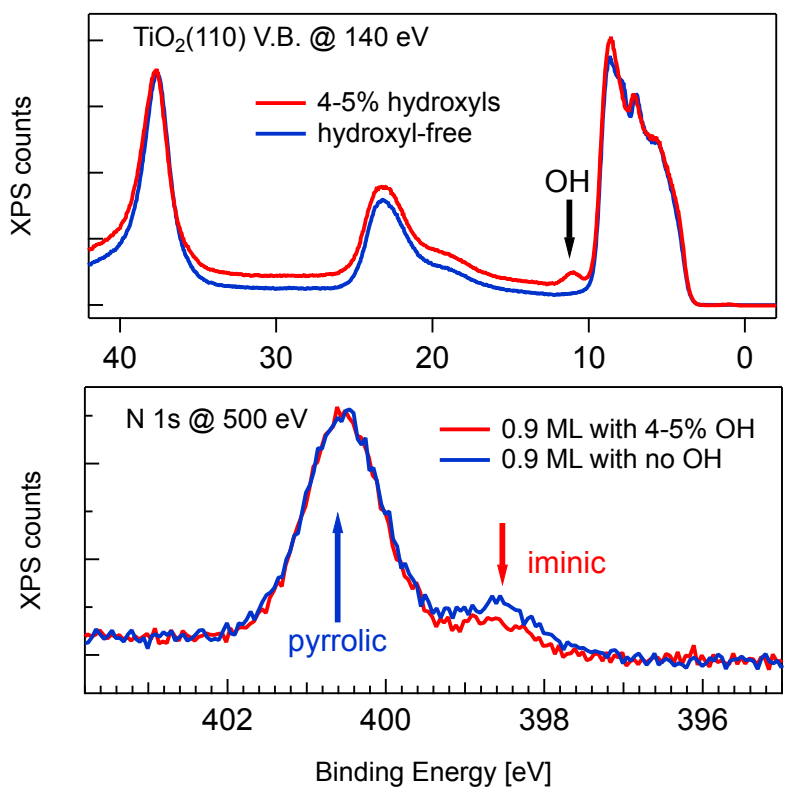

**Figure 1.** Upper panel: Valence Band spectra (hν= 140 eV; ΔE~120 meV) of a slightly hydroxylated (red line) and a hydroxyl-free surface (blue line) before deposition; the hydroxyl peak is marked by an arrow. Lower panel: photoemission spectra of the N 1s peak (hν= 500 eV; ΔE~160 meV) for 0.9 ML of 2H-TPP deposited at RT on the same surfaces of the upper panel; the arrows indicate the pyrrolic and iminic components of molecules in contact with the surface.

In order to clarify the thermodynamics of the self-metalation transition, we have recorded in real time during annealing the high resolution spectra of N 1s starting from a fully hydrogenated (>90%) 4H-TPP contact layer. In Fig. 2, consecutive photoemission spectra are plotted as a function of the substrate temperature. This 2D representation makes evident the decrease of the initially dominating pyrrolic peak and the emergence of a new peak at a binding energy slightly higher than the iminic one, corresponding to incorporation of a Ti atom in the porphyrin macrocycle.

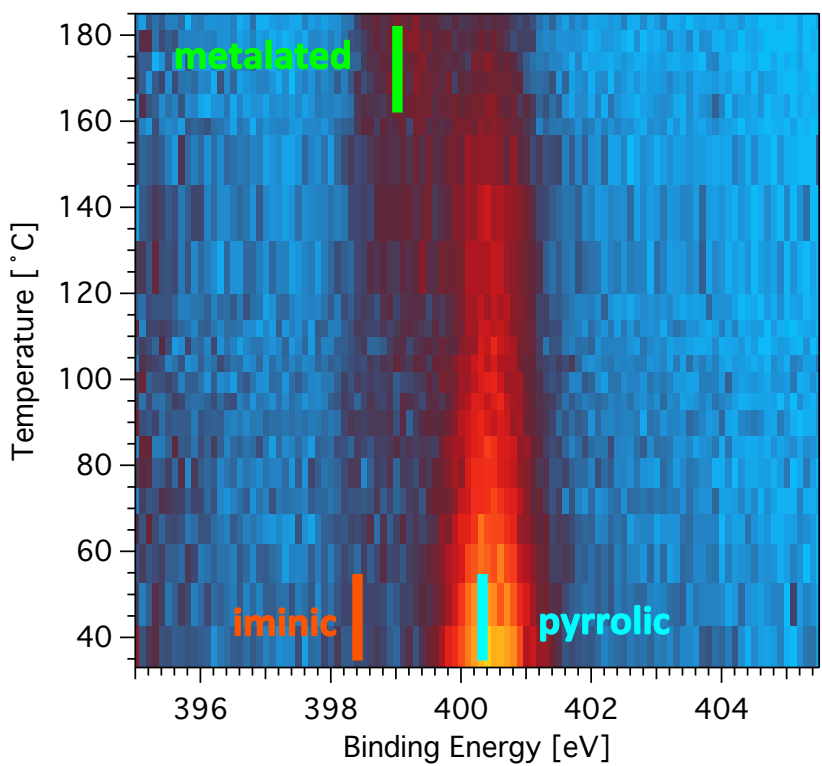

**Figure 2.** 2D intensity map of consecutive N 1s spectra (hν= 500 eV; ΔE~160 meV) taken during substrate heating for ~0.9 ML of 2H-TPP deposited at low temperature, where molecules are almost fully hydrogenated (4H-TPP). The substrate temperature is reported on the left axis. The binding energy of the pyrrolic, iminic and metalated components are also marked.

For comparison, we also measured the N 1s peak after Ti deposition on a 0.8 ML porphyrin layer at room temperature, which yields a peak of NTi at the same binding energy of that obtained by self-metalation (see Fig. 3). The self-metalation approach has the advantage of converting all the molecules in contact with the substrate yielding a layer of purely metalated (100%) porphyrins, as can be appreciated in Fig. 3 from the direct comparison of the N 1s peaks before and after annealing 1 ML, as normalized to the Ti 3p peak. On the contrary, the metalation by Ti deposition cannot convert all the molecules of a compact layer, as previously observed for the case of Ni deposition:[25,40] the deposited Ti atoms soon start to aggregate into metallic clusters due to the limited metal diffusion when the surface is fully covered by molecules. The monolayer self-metalation is completed at ~200°C without further changes in photoemission spectra up to 300°C, however the onset of the chemical reaction takes place at 100°C. We noticed a coverage dependence of the self-metalation temperature, which is typically lower at low coverage (up to 50°C lower for the full metalation with respect to a saturated monolayer, as shown in Fig. 3). This effect is possibly related to the mechanism of Ti incorporation, which however eluded our investigation. Taking into account the high reactivity of Ti atoms, the low temperature value we found for the onset of self-metalation asks for special caution when studying other types of metallo-porphyrins on titanium dioxide, because they might be susceptible to metal exchange (trans-metalation) upon thermal treatment. Remarkably, annealing porphyrin films on titanium dioxide to 100-200°C is a common practice found in several fundamental studies with the aim of desorbing second layer molecules and/or improving film ordering (e.g. Refs [25,40,41]). However, this protocol may jeopardize the interpretation of the results unless the chemical nature of the annealed porphyrin layer is unequivocally assessed by complementary studies. Most importantly, this aspect should be taken in consideration also when designing porphyrin-titania DSSCs because the onset temperature for Ti metalation is very close to that of solar cells during operation conditions. Apart from details of installation and construction geometry, the working temperature of a photovoltaic flat panel ultimately depends on the solar irradiance G and the ambient temperature $T_{air}$.[42] Assuming G~1000 Wm$^{-2}$ and $T_{air}$~40°C (which are not infrequent in a summer day also in most of mediterranean countries and U.S.A. at the sea level[43]) a photovoltaic module can reach 80°C.[44]

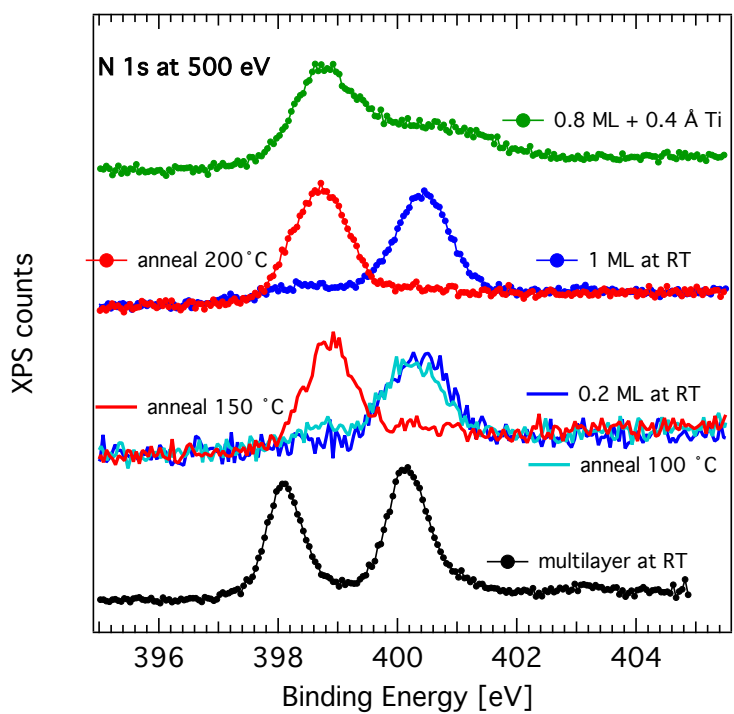

**Figure 3.** From bottom to top we show a comparison among the N 1s photoemission spectra of: a multilayer of 2H-TPP (where the iminic and pyrrolic peak have about the same area); 0.2 ML of 2H-TPP before (blue line) and after annealing to 100°C (magenta line) and 150°C (red line), these three spectra are shown to scale after normalization to the Ti 3p peak for a quantitative comparison of the peak areas; 1 ML of 2H-TPP before (blue markers) and after (red markers) annealing to 200°C (these two spectra are shown to scale after normalization to the Ti 3p peak for a quantitative comparison of the peak areas); a 0.8 ML film after RT deposition of 0.4 Å Ti (displaying partial metalation) that corresponds to an approximate 10:1 atoms per molecule ratio.

Concerning the mechanism of metal incorporation, the capture of Ti atoms detaching from step edges seems unlikely, since we have previously shown by Ti deposition that the compact molecular overlayer inhibits the surface diffusion of free adatoms. We speculate that the discrepancy between the limited mobility of Ti adatoms and the higher reactivity of the substrate Ti atoms is due to the different oxidation states, where the substrate can provide titanium atoms that can readily saturate the nitrogen ligands with minimal change of the Ti coordination number. From XPS measurements, we can also discard a possible capture of interstitials or otherwise located charged Ti atoms. The Ti 2p spectra shown in the upper panel of Fig. 4 indicate the absence of significant variations in the low binding energy tail of the main Ti$_{3/2}$ peak (at 459.1 eV, corresponding to a formal oxidation state +4), where the Ti atoms bearing the excess of charge due to substrate reduction (formal oxidation state +3) would eventually appear (~457 eV). In fact, in this region one would also expect the contribution from titanyl-porphyrin (at 457.8), as recently demonstrated by an experiment of porphyrin metalation through Ti deposition and oxidation on a silver substrate.[45] On the contrary, we observed a significant increase of the Ti 2p$_{3/2}$ low energy tail at 457-458 eV when Ti atoms are deposited on a 2H-TPP layer at RT, as shown in the central panel of Fig. 4. This indicates a mechanism of metal incorporation different from that of self-metalation, thus yielding different metalated porphyrin species, namely physisorbed TiO-TPP species. As a consequence, the Ti atoms incorporated into the porphyrin macrocycle by self-metalation should have about the same oxidation state of the substrate atoms. We may conclude that the porphyrin metal center is likely coordinated to the O$_{br}$ rows underneath, which suggests the formation of a strongly bound Titanium(IV)-TPP species. Finally, from sample to sample and/or at coverage lower than 1 ML, we seldom observed minor variations of the Ti 2p$_{3/2}$ low energy tail upon self metalation, which we ascribe to the formation of either undercoordinated metalated species or to physisorbed titanyl-porphyrins, TiO-TPP. In this regard, we observed minor distortions of the Ti 2p peak in the molecular films upon RHEED illumination, whose diffraction pattern eventually degraded in a few minutes.

In any case, we can exclude any variation of the pristine excess of charge on the substrate Ti atoms upon porphyrin deposition and annealing, which would be otherwise manifested in the valence band. In fact, the excess of charge on the surface, independently on the specific injection mechanism,[45] gives rise to a characteristic peak in the gap (the so called Defect State that has a Ti 3d character), which is correlated to the substrate population of nominal Ti$^{3+}$ atoms. The valence band spectra in the lower panel of Fig. 4 do not show significant intensity variation of the DS peak at 0.9 eV upon molecular deposition at RT and further annealing to 300°C, apart from a small attenuation due to the molecular overlayer, hence the substrate charge distribution is unchanged. Interestingly, the highest occupied molecular orbital, HOMO, of the as deposited layer (4H-TPP) is shifted to higher binding energy with respect to peak measured in a multilayer (2H-TPP). Upon annealing, the HOMO of the metalated layer shifts back to an energy slightly lower than that of the 2H-TPP multilayer. These evidences also support a molecular interaction with the substrate stronger than a simple physisorption.

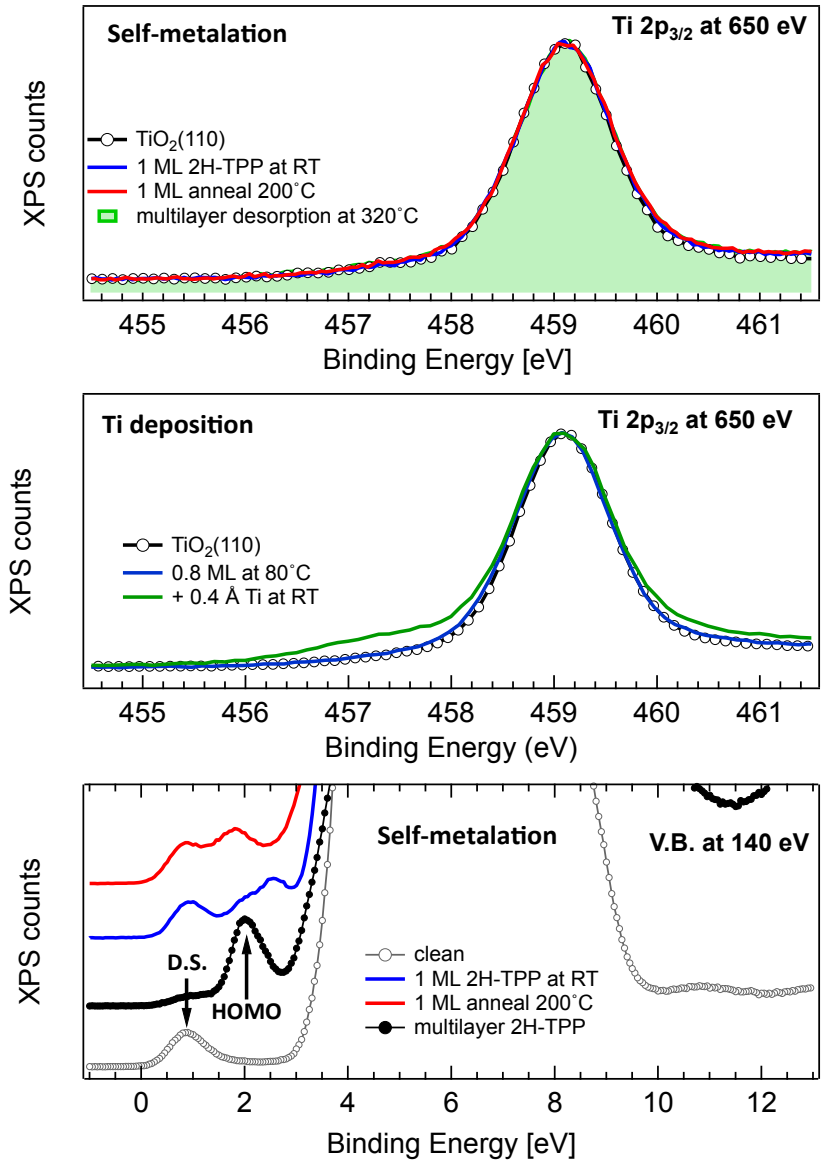

**Figure 4**. Upper panel: normalized photoemission spectra of the Ti $2p_{3/2}$ core level (hν= 650 eV; ΔE~200 meV) for a clean TiO2(110) surface, after deposition of ~1 ML of 2H-TPP at room temperature, and after annealing the latter one to 200°C. For comparison, the spectrum measured on a monolayer obtained by desorption of a multilayer at 320°C is also shown (shaded area). Central panel: normalized photoemission spectra of the Ti $2p_{3/2}$ core level for or a clean TiO2(110) surface, after deposition of a 0.8 ML film (annealed to 80°C to improve molecular ordering; the minor variation of the Ti 2p tail is originated by a small radiation damage due to the short RHEED check performed on this specific film) and after deposition of 0.4 Å of Ti at RT. Lower Panel: photoemission spectra of the valence band (hν= 140 eV; ΔE~120 meV) measured on the same substrate and films of the upper panel. The position of the Defect State peak is indicated by an arrow. The spectra are dominated by the O 2p band (from 3.5 to 9 eV), while the HOMO of the 2H-TPP multilayer is observed at ~2 eV.

**Phase symmetry by RHEED.** In order to determine accurately the phase diagram of the porphyrin monolayer (which corresponds to the most relevant interfacial layer of a real device), we completed our spectroscopy investigation with both surface diffraction and microscopy and compared the energetics of the different phases with the support of ab initio calculations. By means of reflection high energy electron diffraction (RHEED), we have monitored the evolution of the porphyrin superlattice throughout the self-metalation reaction. In order to prevent excessive radiation damage, we illuminated the surface only for a few seconds every 5-10 min (depending on the deposition rate) corresponding to representative deposition stages. Although at low coverage no island formation has been observed, a new diffraction pattern clearly emerges close to the completion of the first layer at room temperature. The new superlattice corresponds to a commensurate $\begin{pmatrix} 2 & 2 \\ 4 & -1 \end{pmatrix}$ symmetry phase, hereafter called *oblique*-(2x4), as shown in the drawings of Fig. 4. Remarkably, the same symmetry phase has been also reported by STM in molecular patches at the monolayer coverage of Zn-TPP.[48] In order to follow the evolution of the *oblique*-(2x4) across the self-metalation, we have monitored by RHEED the superlattice diffraction features along the [1-11] direction, where fractional spots with a characteristic five-fold periodicity can be observed (see pictures in Fig. 5). The faint spots observed at RT become sharper as the substrate is heated step by step up to 300°C, well beyond the full self-metalation of the monolayer, without any change in the angular distribution of the diffracted intensity. Hence, apart from a better ordering due to the mobility increase, we can conclude that the same symmetry phase is observed for porphyrin in its three chemical states, namely, when 2H-TPP and 4H-TPP species are coexisting, as well as after the formation of a 100% pure layer of metalated TPP.

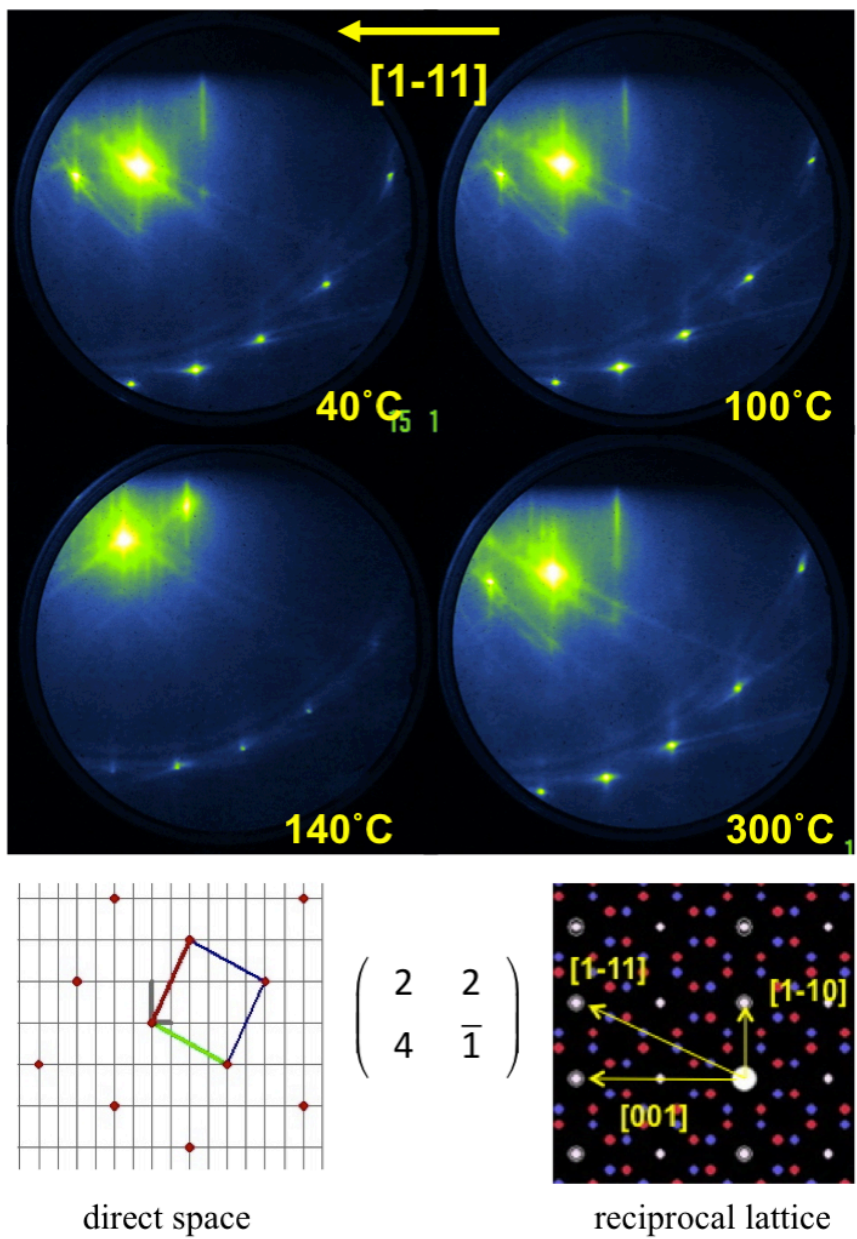

**Figure 5.** RHEED patterns measured at monolayer coverage after annealing at consecutively higher temperatures. Each pattern has been measured after the temperature decreased to <50-60°C in order to have a comparable Debye-Waller attenuation. The surface has been oriented in order to highlight the fractional spots along the [1-11] direction. The corresponding 2D superlattice in the direct and reciprocal space (for the two domains) are shown at the bottom.[47]

**Molecular arrangement by STM topography.** These findings, concerning the very long range order (RHEED transfer width >1000 nm) as averaged on a large area (8x1 mm$^2$) by the grazing footprint of the electron beam, are also consistent with the molecular ordering observed locally by STM. At low coverage, single molecules are clearly recognized by their characteristic saddle-shape appearance and they are seen to adsorb atop the $O_{br}$ rows due to the hydrogen bonding of their original pyrrolic groups (see top-left panel in Fig. 6).[16] In this regime of growth, porphyrins do not display either preferential aggregation into islands (as due to intermolecular attraction) or regular spacing (as due to intermolecular repulsion). As the coverage increases and diffusing molecules along adjacent $O_{br}$ rows enter into contact, the phenyl rings of adjacent molecules interact and start to arrange the molecules in an ordered scheme corresponding to the *oblique*-(2x4) phase (see top-right panel of Fig. 6), which displays the highest density packing ($\rho = 0.52$ nm$^{-2}$) we observe, up to the completion of the monolayer. In this regime of growth (0.4-0.6 ML) we cannot exclude the local formation of domains with $\begin{pmatrix} 2 & -2 \\ 5 & -1 \end{pmatrix}$ symmetry, hereafter called *oblique*-(2x5), corresponding to a lower density phase ($\rho = 0.43$ nm$^{-2}$). The latter phase has been also claimed for the Zn-TPP monolayer phase, as obtained upon annealing to 150°C a film in excess of one monolayer.[41] However, we remark that after annealing to high temperature either a carefully calibrated layer (from 0.8 to 1 ML) or a film in excess of 1 monolayer (as in the case of the bottom panel in Fig. 6), we always obtain the same *oblique*-(2x4) symmetry phase, which corresponds to the saturation monolayer coverage.

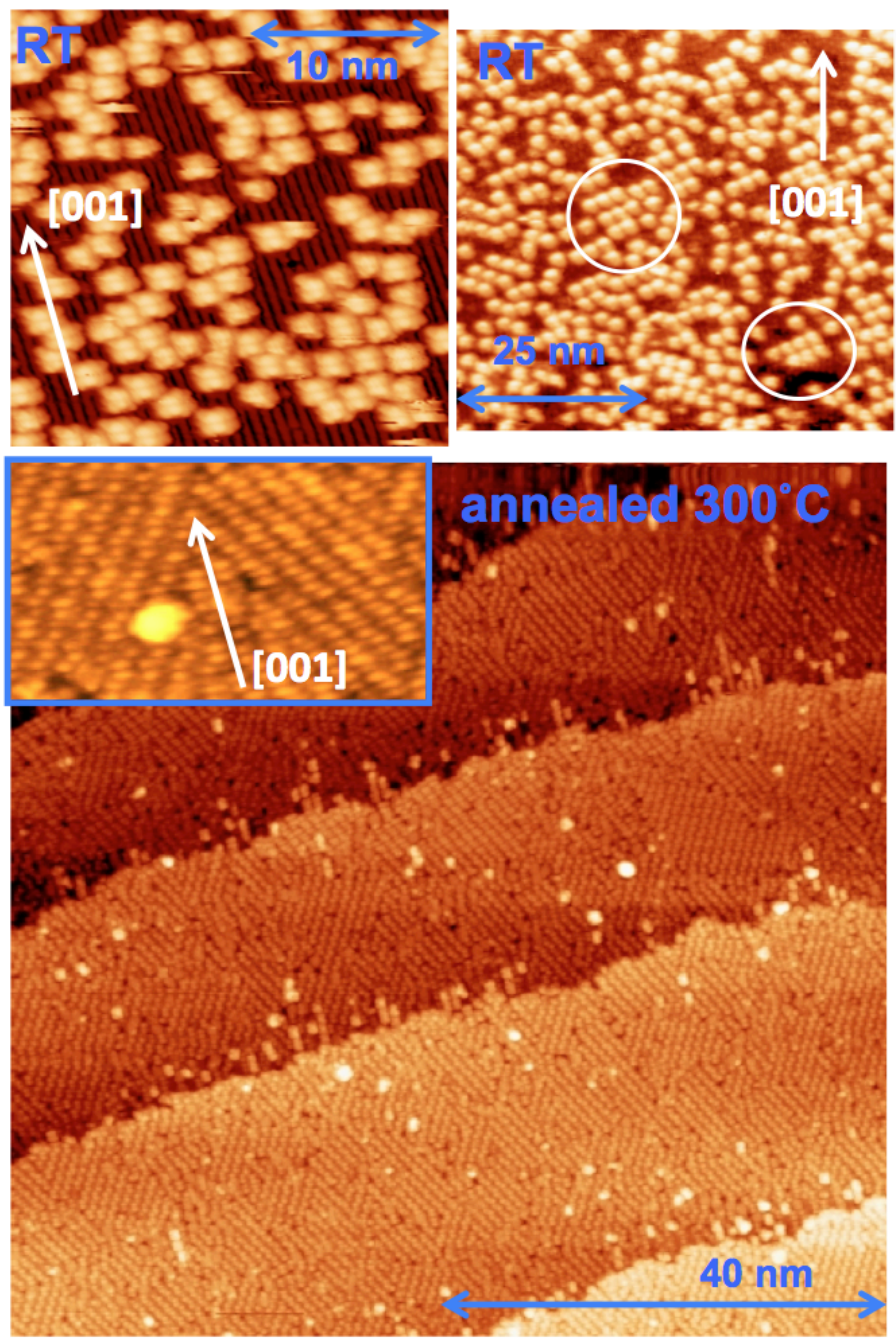

**Figure 6.** Top: STM images of 2H-TPP (and 4H-TPP) films of different coverage. Left: ~0.2 ML at RT, free-base molecules randomly distributed atop the $O_{br}$ rows (dark lines in the background) with homogeneous azimuthal orientation and saddle-shape appearance (-1.4V, 60pA). Right: ~0.5 ML at RT, onset of an intermolecular spatial correlation transverse to the rows locally showing small patches of 10-12 molecules with *oblique*-(2x4) symmetry (highlighted by circles). Bottom: large scale of 1 ML obtained by thermal desorption at ~300°C of a bi-layer (+2.5V, 100pA). All terraces are covered by a close compact *oblique*-(2x4) phase entirely composed of metalated molecules; inset shows the saddle-shape of the metalated molecules.

**Porphyrin adsorption geometry by DFT.** In order to draw an atomistic model of the *oblique*-(2x4) phase, we performed DFT calculations with the Quantum ESPRESSO package[31] adopting Grimme dispersion corrections[33] and repeated slab models with PBE-D theoretical lattice constants of the substrate.[35] As shown in our former study,[16] we had consistently found that the equilibrium adsorption site for both free-base porphyrin 2H-TPP and its doubly hydrogenated counterpart 4H-TPP is the bridge site between to oxygen atoms on the $O_{br}$ rows. This configuration is driven by the formation of hydrogen bonds between the two original pyrrolic groups NH and the corresponding $O_{br}$ beneath. In fact, according to our simulations, the two types of molecule cannot be discriminated by STM and their topographic contrast is dominated by an enhanced saddle-shape of the macrocycle appearing as a nodal plane transverse to the rows (see right panels in Fig. 6). The bare incorporation of a Ti atom in the macrocycle upon dehydrogenation of the pyrrolic groups does not change the geometric scenario. The most favored molecular configuration for an isolated molecule on a clean and stoichiometric surface corresponds to the replacement of the two hydrogen bonds NH•••$O_{br}$ with two NTi•••$O_{br}$ bonds on the row of protruding oxygen atoms. The molecule, hereafter called $TiO_2$-TPP, remains on a bridge site on the $O_{br}$ rows of protruding oxygen atoms with minor relaxation changes of the molecular backbone. The corresponding STM simulations do not show significant differences between the metalated and hydrogenated species, in full agreement with our STM images, where the molecules display the same saddle-shape appearance across the self-metalation reaction (compare the molecular contrast in the top and bottom panels of Fig. 7, corresponding to free-base and metalated porphyrins, respectively).

For a better understanding of the local mechanism driving the adsorption geometry, we also tested the adsorption atop a single $O_{br}$ atom (with a double bond), as recently proposed for the equilibrium adsorption configuration of Ni-TPP and Zn-TPP,[40,41] and alternative molecular ordering. In the case of an isolated molecule, we find the bridge site to be favored by 0.07 eV with respect to the on-top one. When considering correlated molecules, i.e. the ordered domains, the energy balance is inverted for the *oblique*-(2x4) phase, where the on-top site is now energetically favored by the same amount of 0.07 eV. In contrast, the bridge site remains the preferred one for an *oblique*-(2x5) phase with respect to the on-top by ~0.1 eV. The adsorption geometry of Ti-metalated TPP seems to be the result of two contrasting effects: on one side, the molecular backbone prefers the bridge site to optimize the network of H-bonds with the surface, on the other side, Ti atoms prefer to form a double Ti=O bond as in the TiO-TPP molecule. H-bonds prevail over the Ti=O bond in both the isolated and low density *oblique*-(2x5) phases. On the contrary, the denser *oblique*-(2x4) arrangement weakens the phenyl H-bond network through a small azimuthal reorientation, so that the Ti=O bond prevails. However, since the metalated phase is formed starting from the already formed *oblique*-(2x4) phase, mostly made of 4H-TPP, a collective movement of the molecules to change the adsorption site is unlikely and we can conclude that a purely $TiO_2$-TPP film is formed upon Ti incorporation, in agreement with the absence of any change of both the photoemission spectra of Ti 2p and the RHEED patterns across the self-metalation reaction. Contrary to previous expectations,[27] the porphyrin self-metalation on $TiO_2$(110) does not proceed by extraction of titanyl, $TiO^{2+}$, radicals to form TiO-TPP free molecules on the surface. Rather, 2H/4H-TPP molecules are trapped at $O_{br}$ rows, where they can incorporate a bare Ti atom without breaking their chemical bond to the substrate.

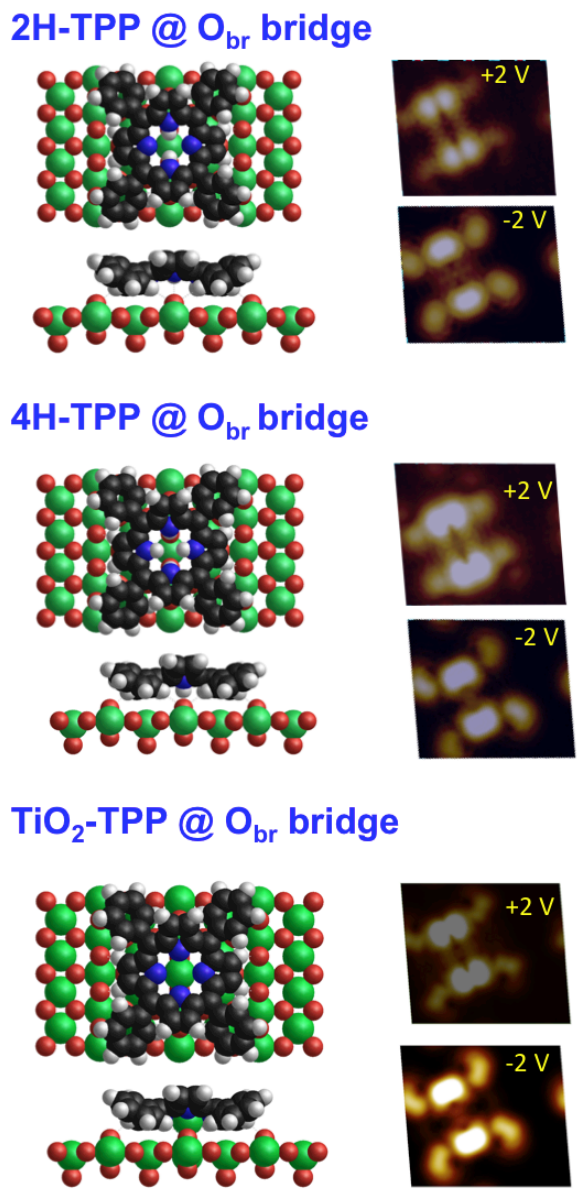

**Figure 7.** A sideview of the adsorption geometry of 2H-TPP, 4H-TPP and TiO$_2$-TPP is shown on the left, highlighting the interaction of the NH and NTi groups with the O$_{br}$ rows underneath. The corresponding simulated STM images are shown on the right for both positive and negative sample bias (also see Ref. [16] for a comparison with high resolution STM images).

In order to understand the energetics of the ordered phases, we calculated the adsorption energy (E$_{ads}$) for all of the four possible chemical states, 2H-TPP, 4H-TPP, TiO-TPP and TiO$_2$-TPP in a few periodic arrangements of different densities. The equilibrium configuration of each phase is obtained starting from the configuration found for the isolated ones and leads to minor relaxations due to the phenyls intermolecular coupling. The results are summarized in Fig. 8 for the *oblique*-(2x4) and *oblique*-(2x5) phases that are representative of two different densities. Molecules in the *oblique*-(2x4) phase display a small azimuthal reorientation of ~8° due to the aforementioned phenyl interaction among adjacent molecules, whereas the in-plane molecular axis remain perfectly aligned to the high symmetry direction in the *oblique*-(2x5) phase. We find that the molecular adsorption energy decreases as the density increases for all chemical species, i.e. molecules in the *oblique*-(2x4) phase are less bound to the substrate than in the *oblique*-(2x5) one, (see bottom-left panel of Fig. 8) which are also less bound than isolated molecules. The difference is very small for the case of 2H-TPP and it is largest for the metalated molecules. Increasing the film density is however convenient when we consider the specific adsorption energy $\varepsilon_{ads}$ obtained by weighting the adsorption energy with the film density, $\varepsilon_{ads} = E_{ads} \times \rho$, (see bottom-right panel of Fig. 8). In this case, the energy gain for increasing the phase density is maximum for 4H-TPP (0.35 eV) and minimum for the metalated species. In particular, the *oblique*-(2x5) phase on the bridge site becomes slightly favored by 0.09 eV with respect to the *oblique*-(2x4) one. However, this small energy difference is not sufficient to change the already established *oblique*-(2x4) high density ordering of the hydrogenated porphyrin film when it undergoes the metalation reaction, as experimentally observed from room temperature up to 300°C.

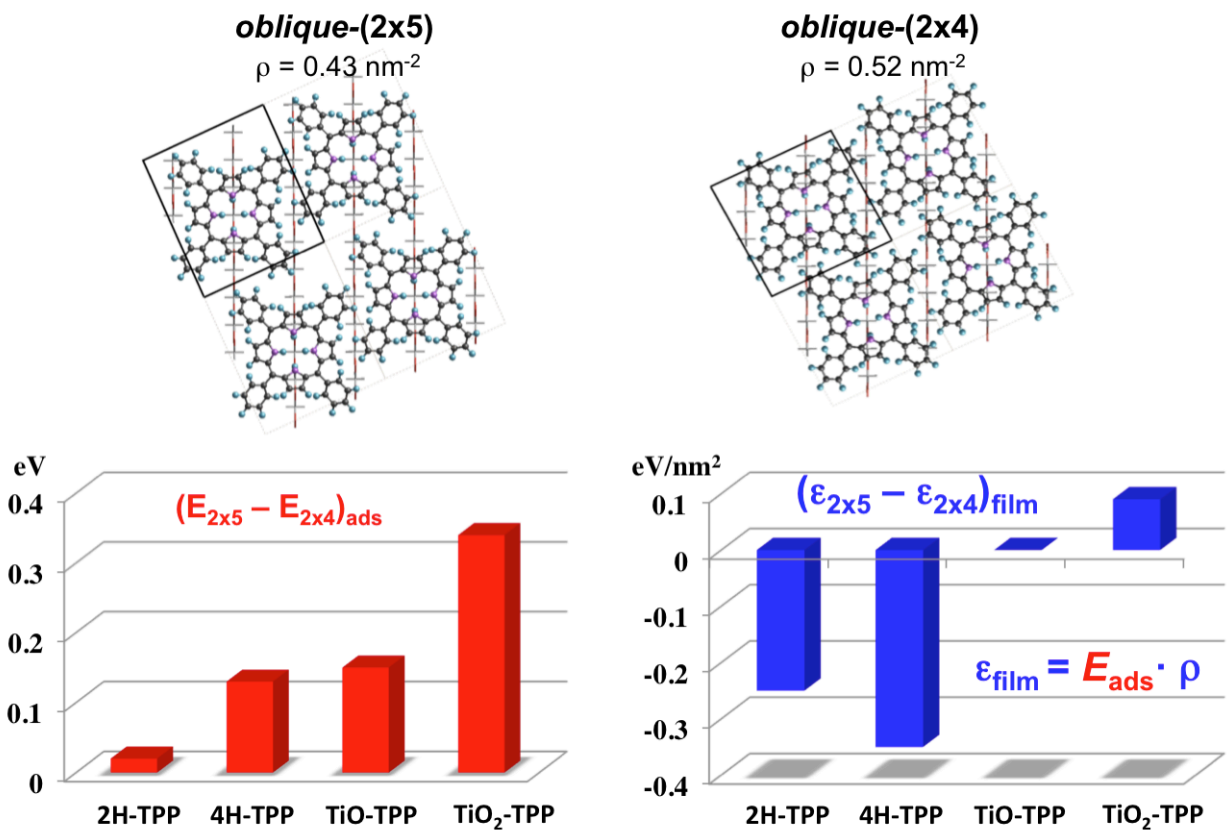

**Figure 8.** Upper panels: topview of the relaxed structure obtained for the *oblique*-(2x5) and *oblique*-(2x4) phases (right and left, respectively); the corresponding superlattice unit cell is also indicated; all molecules are on the $O_{br}$ rows, on bridge sites for 2H-, 4H-TPP and $TiO_2$-TPP, on top sites for TiO-TPP. Lower panels: comparison of the molecular $E_{ads}$ and film $\varepsilon_{ads}$ adsorption energies (left and right, respectively) for the four porphyrin species in the two different phases; positive $E_{ads}$ difference (left panel in red) favors the molecular adhesion in the *oblique*-(2x5); negative $\varepsilon_{ads}$ difference (right panel in blue) favors the overall stability of the *oblique*-(2x4) phase.

CONCLUSIONS

In conclusion, we have demonstrated by a combination of spatially averaged and local probes, that free-base tetra-phenyl-porphyrins form a long range ordered commensurate phase at the monolayer coverage on rutile TiO$_2$(110), which is fully converted into a film of Titanium(IV)-porphyrin, TiO$_2$-TPP, by a self-metalation reaction without changing the molecular ordering. By DFT simulations, we have shown that the observed phase with $\begin{pmatrix} 2 & 2 \\ 4 & -1 \end{pmatrix}$ symmetry, corresponding to the most dense monolayer porphyrin packing on TiO$_2$(110), is formed by molecules adsorbed on a bridge site between two O$_{br}$ atoms for porphyrins in all of the three chemical states we found, namely 2H-TPP, 4H-TPP and TiO$_2$-TPP. Thanks to the strong porphyrin bonding to the substrate, this phase persists from room temperature up to 300°C, thus making this interface technological relevant for possible applications in photocatalysis. Most importantly, the self-metalation reaction sets in at a critical temperature as low as ~100°C. Hence, in order to prevent metal exchange processes with the highly reactive Ti atoms, much attention must be payed to the control of temperature when other metal-porphyrins are brought into contact with titania surfaces, as in the case of photovoltaic devices, which can easily reach 80°C in working conditions.

AUTHOR INFORMATION

**Corresponding Authors**

*floreano@iom.cnr.it, andrea.vittadini@unipd.it

**Present Addresses**

# Department of Applied Physics, Columbia University, New York, NY 10027, USA.

**Author Contributions**

The manuscript was written through contributions of all authors. All authors have given approval to the final version of the manuscript.

ACKNOWLEDGMENT
We acknowledge support from the MIUR of Italy through PRIN project DESCARTES (n. 2010BNZ3F2), from the Spanish Government (Grant MAT2013-46593-C6-4-P) and from the Basque Department of Education (Grant IT-621-13). We acknowledge the CINECA award under the ISCRA initiative, for the availability of high performance computing resources and support. M.A. also thanks the financial support of MPC.

ABBREVIATIONS

XPS, X-ray Photoemission Spectroscopy; STM, Scanning Tunneling Microscopy; RHEED, RHEED, Reflection High Energy Electron Diffraction; DFT, Density Functional Theory; TPP, Tetra-Phenyl-Porphyrin.

TOC Graphic

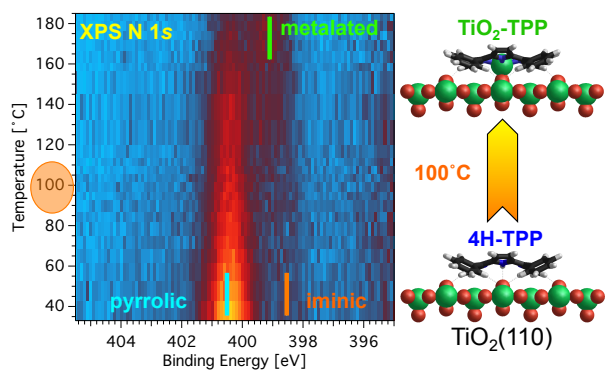